\documentclass[graybox]{svmult}
\usepackage{mathptmx}       % selects Times Roman as basic font
\usepackage{helvet}         % selects Helvetica as sans-serif font
\usepackage{courier}        % selects Courier as typewriter font
\usepackage{type1cm}        % activate if the above 3 fonts are
\usepackage{makeidx}         % allows index generation
\usepackage{graphicx}        % standard LaTeX graphics tool
\usepackage{multicol}        % used for the two-column index
\usepackage[bottom]{footmisc}% places foot\cite{Rend92}notes at page bottom

\usepackage{amsfonts}
\usepackage{amsmath}
\usepackage{amssymb}
\usepackage{bm}
\usepackage{hyperref}
\usepackage{color}

\allowdisplaybreaks
\newcommand{\ua}{^{\alpha}}
\newcommand{\ub}{^{\beta}}
\newcommand{\la}{_{\alpha}}

\newcommand{\ubar}{{\bar u}}

\newcommand{\ret}{{\text{ret}}}
\newcommand{\s}{{\text{S}}}
\newcommand{\R}{{\text{R}}}
\newcommand{\MU}{{m_1}}
\newcommand{\m}{{m_2}}

\newcommand{\beq}{\begin{equation}}
\newcommand{\eeq}{\end{equation}}

\newcommand{\C}{\mathbb{C}}
\newcommand{\ud}{\mathrm{d}}

\newcommand{\ab}{^{\alpha\beta}}
\newcommand{\lab}{_{\alpha\beta}}

\newcommand{\uuhh}{{\bar{u}^\alpha \bar{u}^\beta \hat{h}^{\text{R}}_{\alpha\beta}}}

\newcommand{\chsq}{{\chi^2}}

\newcommand{\uT}{u^T}
\newcommand{\SF}{{\text{SF}}}
\newcommand{\hR}{{h^\text{R}_{\alpha\beta}}}
\newcommand{\hhR}{{\hat{h}^\text{R}_{\alpha\beta}}}

\makeindex             % used for the subject index
\begin{document}

\title*{High-accuracy comparison between the post-Newtonian and self-force dynamics \\ of black-hole binaries}

\titlerunning{High-accuracy comparison between the PN and SF dynamics of black-hole binaries}

\author{Luc Blanchet, Steven Detweiler, Alexandre Le Tiec and Bernard F. Whiting}

\institute{
Luc Blanchet \at
${\mathcal{G}}{\R}\varepsilon{\C}{\mathcal{O}}$, Institut d'Astrophysique de Paris --- C.N.R.S. \& Universit\'e Pierre et Marie Curie, 98$^{\mathrm{bis}}$ boulevard Arago, 75014 Paris, France. \email{blanchet@iap.fr}
\and Steven Detweiler \at
Institute for Fundamental Theory, Department of Physics, University of Florida, Gainesville, FL 32611-8440, USA. \email{det@phys.ufl.edu}
\and Alexandre Le Tiec \at
${\mathcal{G}}{\R}\varepsilon{\C}{\mathcal{O}}$, Institut d'Astrophysique de Paris --- C.N.R.S. \& Universit\'e Pierre et Marie Curie, 98$^{\mathrm{bis}}$ boulevard Arago, 75014 Paris, France. \email{letiec@iap.fr}
\and Bernard F. Whiting \at
Institute for Fundamental Theory, Department of Physics, University of Florida, Gainesville, FL 32611-8440, USA. \email{bernard@phys.ufl.edu}}
\maketitle

\vspace{-0.5cm}
\abstract{The relativistic motion of a compact binary system moving in circular orbit is investigated using the post-Newtonian (PN) approximation and the perturbative self-force (SF) formalism. A particular gauge-invariant observable quantity is computed as a function of the binary's orbital frequency. The conservative effect induced by the gravitational SF is obtained numerically with high precision, and compared to the PN prediction developed to high order. The PN calculation involves the computation of the 3PN regularized metric at the location of the particle. Its divergent self-field is regularized by means of dimensional regularization. The poles $\propto (d-3)^{-1}$ which occur within dimensional regularization at the 3PN order disappear from the final gauge-invariant result. The leading 4PN and next-to-leading 5PN conservative logarithmic contributions originating from gravitational-wave tails are also obtained. Making use of these exact PN results, some previously unknown PN coefficients are measured up to the very high 7PN order by fitting to the numerical self-force data. Using just the 2PN and new logarithmic terms, the value of the 3PN coefficient is also confirmed numerically with very high precision. The consistency of this cross-cultural comparison provides a crucial test of the very different regularization methods used in both SF and PN formalisms, and illustrates the complementarity of these approximation schemes when modelling compact binary systems.} 

\section{Introduction and motivation}\label{sec1}

For the gravitational wave observatories LIGO/Virgo/GEO on Earth and LISA in space, routine identification of inspiralling compact binaries (binary systems composed of neutron stars and/or black holes) will require high-accuracy predictions from general relativity theory \cite{Th300,3mn}. Providing such predictions represents a formidable task that can be addressed using approximation schemes in general relativity. The two main approximation schemes available are: (i) the \textit{post-Newtonian} expansion, well suited to describe the inspiralling phase of compact binaries in the slow motion and weak field regime independently of the mass ratio, and (ii) the \textit{self-force} approach, based on perturbation theory, which gives an accurate description of extreme mass ratio binaries even in the strong field regime.

The post-Newtonian (PN) templates for compact binary inspiral have been developed to 3.5PN order in the phase \cite{BIJ02,BFIJ02,BDEI04} and 3PN order in the amplitude \cite{BIWW96,BFIS08} (see Blanchet's contribution in this volume).\footnote{As usual the $n$PN order refers to terms equivalent to $(v/c)^{2n}$ beyond Newtonian theory, where $v$ is a typical internal velocity of the material system and $c$ is the speed of light.} These are suitable for the inspiral of two neutron stars in the frequency bandwidth of LIGO and Virgo detectors. For detection of black hole binaries (with higher masses), the construction of template banks either requires the matching of the PN waveform with full numerical simulations for the merger phase and the ringdown of the final black hole \cite{Boyle08,Ajith08}, or using the effective-one-body formalism \cite{BuonD99} (see also Nagar's contribution in this volume).

In a completely different parameter regime, gravitational self-force (SF) analysis \cite{MiSaTa,QuWa,DW03,GW08,Poisson} (see also Poisson's contribution in this volume) is expected to provide templates for extreme mass ratio inspirals (EMRIs) anticipated to be present in the LISA frequency bandwidth. SF analysis is a natural extension of first order perturbation theory, and the latter has a long history of comparisons with post-Newtonian analysis \cite{P93a,CFPS93,P93b,TNaka94,P95,TTS96,TSTS96,PS95}. SF analysis itself, however, is just now mature enough to present some limited comparisons with PN analysis, although it is not yet ready for template generation. 

\begin{figure}
    \includegraphics[width=11.34cm]{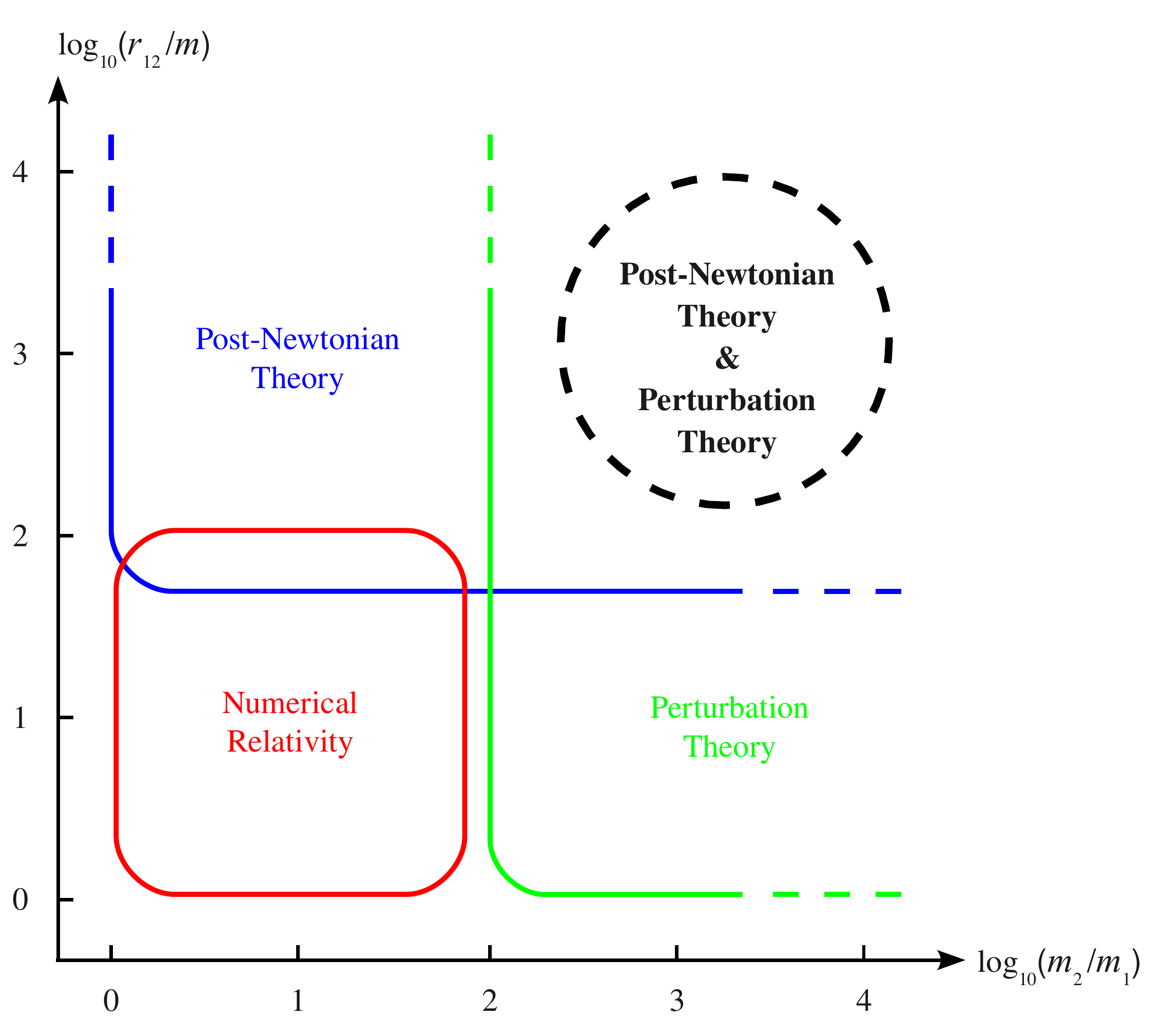}
    \caption{\footnotesize Different analytical approximation schemes and numerical techniques are used to study black hole binaries, depending on the mass ratio $m_1/m_2$ and the orbital velocity $v^2 \sim G m/r_{12}$, where $m = m_1 + m_2$. The post-Newtonian theory and black hole perturbation theory can be compared in the slow motion regime ($v\ll c$ equivalent to $r_{12}\gg G m/c^2$ for circular orbits) of an extreme mass ratio ($m_1 \ll m_2$) binary.}
    \label{methods}
\end{figure}

In recent works \cite{BDLW10a,BDLW10b} (hereinafter referred to as Papers I and II respectively) we performed a high-accuracy comparison between the PN and SF analyses in their common domain of validity, that of the slow motion weak field regime of an extreme mass ratio binary (see illustration of various methods in Fig.~\ref{methods}). The problem was tackled previously by Detweiler \cite{Det08}, who computed numerically within the SF approach a certain gauge invariant quantity (called the redshift observable), and compared it with the 2PN prediction extracted from existing PN results \cite{BFP98}. We then extended this comparison in Papers I-II up to higher PN orders. This required an improvement in the numerical resolution of the SF calculation in order to distinguish more accurately the various contributions of very high PN order terms. However, our primary difficulty has been that the relevant PN results for the metric were previously not available beyond the 2PN order, and had to be carefully derived. We have finally demonstrated an excellent agreement between the SF contribution to the analytical PN result (derived through 3PN order, with inclusion of specific logarithmic terms at 4PN and 5PN orders) and the exact numerical SF result.

In this article we present a summary of the Papers I and II. The plan is as follows: After having introduced in Sec.~\ref{method} the coordinate-invariant relation used to perform the comparison, we describe in Sec.~\ref{regularization} how the divergent self-field of point particles is regularized in both SF and PN formalisms. Sec.~\ref{overview} provides a brief overview of how the SF computation proceeds. Secs.~\ref{3PN} and \ref{logs} present the post-Newtonian computations of the 3PN regularized metric, as well as of the 4PN and 5PN logarithmic contributions. The PN results for the gauge-invariant relation are discussed in Sec.~\ref{PN}. Finally, Sec.~\ref{secVI} is devoted to the comparison of these PN results with the SF numerical data, and the measurement of unknown high order PN coefficients.

\section{The gauge invariant redshift observable}
\label{method}

We consider a system of two (non-spinning) compact objects with masses $m_1$ and $m_2$, and moving on slowly inspiralling quasi-circular orbits. In the PN analysis, let $m_1$ and $m_2$ be arbitrary; in the SF analysis, further assume that $m_1 \ll m_2$. We can then call $m_1$ the ``particle'', and $m_2$ the ``black hole''.

Self-force analysis shows that the dissipative parts of the self-force for a circular orbit are the $t$ and $\varphi$ components. These result in a loss of energy and angular momentum from the small mass at the same precise rate as energy and angular momentum are radiated away \cite{Det08}. In addition, earlier perturbative calculations of energy and angular momentum fluxes \cite{P93a,CFPS93,P93b,TNaka94,P95,TTS96,TSTS96,PS95} for this situation show them to be equivalent to the results of the PN analysis in their common domain of validity. Hence, by invoking an argument of energy and angular momentum balance, we know that the PN results also agree with the dissipative parts of the SF, and further comparison can reveal nothing new.

For our PN-SF comparison, we shall thus neglect the dissipative, radiation-reaction force responsible for the inspiral, and restrict ourselves to the conservative part of the dynamics. In PN theory this means neglecting the dissipative radiation-reaction force at 2.5PN and 3.5PN orders, and considering only the conservative dynamics at the even-parity 1PN, 2PN and 3PN orders. This clean separation between conservative even-parity and dissipative odd-parity PN terms is correct up to 3.5PN order. However such split breaks at 4PN order, since at that approximation arises a contribution of the radiation-reaction force, which originates from gravitational wave tails propagating to infinity \cite{BD88} (this will be further discussed in Sec.~\ref{logs}). In SF theory there is also a clean split between the dissipative and conservative parts of the self-force (see e.g. \cite{Ba09}). This split is particularly transparent for a quasi-circular orbit, where the $r$ component is the only non-vanishing component of the conservative self-force.

Henceforth, the orbits of both masses are assumed to be and to remain circular, because we are ignoring the dissipative radiation-reaction effects. For our comparison we require two physical quantities which are precisely defined in the context of each of our approximation schemes. The orbital frequency $\Omega$ of the circular orbit as measured by a distant observer is one such quantity. The second quantity is defined as follows. 

With circular orbits and no dissipation, the geometry has a helical Killing vector field $k\ua$. A Killing vector is only defined up to an overall constant factor. In our case $k^\alpha$ extends out to a large distance where the geometry is essentially flat. There, 
\begin{equation}\label{killing}
	k\ua\partial\la = \partial_t + \Omega\,\partial_\varphi\,,
\end{equation}
in any natural coordinate system which respects the helical symmetry \cite{SBD08}. We let this equality define the overall constant factor, thereby specifying the Killing vector field uniquely.

An observer moving with the particle $m_1$, while orbiting the black hole $m_2$, would detect no change in the local geometry. Thus the four-velocity $u_1^\alpha$ of the particle is tangent to the Killing vector $k^\alpha$ evaluated at the location of the particle, which we denote by $k_1\ua$. A second physical quantity is then defined as the constant of proportionality, call it $u_1^T$, between these two vectors, namely
\begin{equation}\label{ut_def}
   u_1\ua = u_1^T \,k_1\ua\,.
\end{equation}
The four-velocity of the particle is normalized so that ${(g\lab)}_1 u_1\ua u_1\ub = -1$; ${(g\lab)}_1$ is the \textit{regularized} metric at the particle's location, whereas the metric itself is formally singular at the particle $m_1$ in both PN and SF approaches. 

If we happen to choose a coordinate system such that \eqref{killing} is satisfied everywhere, then in particular $k_1^t=1$, and thus $u_1^T\equiv u_1^t$, the $t$ component of the four-velocity of $m_1$. The Killing vector on the particle is then $k_1\ua=u_1\ua/u_1^t$, and simply reduces to the particle's ordinary post-Newtonian coordinate velocity $v_1\ua/c$. In such a coordinate system, the description of the invariant quantity we are thus considering is
\begin{equation}\label{ut_defX}
    u_1^T \equiv u_1^t = \biggl( - {(g\lab)}_1 \frac{v_1\ua v_1\ub}{c^2} \biggr)^{-1/2} \,.
\end{equation}
It is important to note that this quantity is precisely defined in both PN and SF frameworks, and it does not depend upon the choice of coordinates or upon the choice of perturbative gauge. The quantity $u_1^T$ represents the redshift of light rays emitted from the particle and received on the helical symmetry axis perpendicular to the orbital plane \cite{Det08}; we shall refer to it as the \textit{redshift observable}.

\section{Regularization issues in the SF and PN formalisms}
\label{regularization}

The redshift observable \eqref{ut_defX} depends upon using a valid method of regularization. The regularized metric ${(g\lab)}_1$ is defined with very different prescriptions in the SF and PN approaches. Both analyses require subtle treatment of singular fields at the location of the masses. Subtracting away the infinite part of a field while carefully preserving the part which is desired is always a delicate task.

In the SF prescription, the regularized metric reads
\begin{equation}\label{regSF}
    g\lab^\mathrm{SF}(x)=\bar{g}\lab(x)+h\lab^\mathrm{R}(x)\,,
\end{equation}
where $\bar{g}\lab$ denotes the background Schwarzschild metric of the black hole, and where the ``Regular'' metric perturbation $h\lab^\mathrm{R}$ is smooth in a neighborhood of the particle, and given by the difference
\begin{equation}
h^\R\lab = h^\ret\lab - h^\s\lab
\end{equation}
between the retarded metric perturbation $h^\ret\lab$ and the purely locally determined ``Singular'' field $h^\s\lab$. Following the Detweiler-Whiting prescription \cite{DW03}, a Hadamard expansion of Green's functions in curved spacetime provides an expansion for $h^\s\lab$ \cite{DW03}. In a neighborhood of the particle with a special, locally-inertial coordinate system, $h^\s\lab$ appears as the $m_1/r$ part of the particle's Schwarzschild metric along with its tidal distortion caused by the background geometry of the large black hole. Details of the expansion are given in Sec.~6.1 of \cite{Det05}. The special locally inertial coordinates for a circular geodesic in the Schwarzschild metric are given as functions of the Schwarzschild coordinates in Appendix~B of \cite{DMW03}. (See also Detweiler's and Barack's contributions in this volume.) Since the metric \eqref{regSF} is regular at the particle's position $y_1\ua$, we simply have
\begin{equation}\label{regSF1}
    {(g\lab^\mathrm{SF})}_1=g\lab^\mathrm{SF}(y_1)\,.
\end{equation}
In the perturbative SF analysis we are only working through first order in the mass ratio $q\equiv m_1/m_2$, and at that level of approximation $h^\R\lab = \mathcal{O}(q)$. Then $u_1^T$ can be computed accurately to the same perturbative order and compares well with the post-Newtonian result to 2PN order \cite{Det08}. The regularized 2PN metric is known \cite{BFP98}, and therefore the comparison is straightforward.

In the PN prescription on the other hand, one first computes the metric
$g\lab^\mathrm{PN}(\mathbf{x},t)$ using an iterative PN procedure at any field
point outside the particle, in a coordinate system $x\ua=\{ct,x^i\}$. That
metric is generated by the two particles, and includes both regular and
singular contributions around each particle. Such iterative PN calculation is
a very long and intricate procedure up to say 3PN, at which order it will be
partly based on existing computations of the 3PN equations of motion using
Hadamard \cite{BFeom} and dimensional \cite{BDE04} regularizations. Then we
compute the regularized metric at the location of the particle by taking the
limit when $\mathbf{x}\rightarrow\mathbf{y}_1(t)$, where $\mathbf{y}_1(t)$ is
the particle's trajectory. In 3 spatial dimensions, that limit is singular. In
order to treat the infinite part of the field, we extend the computation in
$d$ spatial dimensions, following the prescription of \textit{dimensional
  regularization} \cite{tHooft,Bollini}, which is based on an analytic
continuation (AC) in the dimension $d$ viewed as a complex number. We do
not use Hadamard's regularization which found its limit at 3PN order.
Considering the analytic continuation in a neighborhood of
$\varepsilon\equiv d-3 \rightarrow 0$, we define
\begin{align}\label{gPNdr}
    {(g\lab^\mathrm{PN})}_1 = \mathop{\mathrm{AC}}_{\varepsilon\rightarrow 0}\,\Bigl[\lim_{\mathbf{x}\rightarrow\mathbf{y}_1} g\lab^\mathrm{PN}(\mathbf{x},t)\Bigr]\,.
\end{align}
The limit $\varepsilon\rightarrow 0$ does not exist in general due to the presence of poles $\propto\varepsilon^{-1}$ occuring at the 3PN order; we thus do not take the strict limit $\varepsilon \rightarrow 0$ but compute the singular Laurent expansion when $\varepsilon\rightarrow 0$. At 3PN order the result takes the schematic form
\begin{equation}\label{g1dr}
   {(g\lab^\mathrm{PN})}_1  = \frac{1}{\varepsilon} g^{(-1)}\lab(\mathbf{y}_1,t) + g^{(0)}\lab(\mathbf{y}_1,t) + \mathcal{O}(\varepsilon) \,,
\end{equation}
where $g^{(-1)}\lab(\mathbf{y}_1,t)$ denotes the pole part, which is purely of 3PN order, and $g^{(0)}\lab(\mathbf{y}_1,t)$ is the finite part. At higher PN orders we expect the presence of multipole poles $\propto\varepsilon^{-n}$. As we shall see, the (simple) poles at 3PN order will disappear from the final gauge invariant relationship $u_1^T(\Omega)$. In fact the occurence of poles at the 3PN order is specific to the use of the harmonic gauge condition. Previous work on the 3PN equations of motion of point particle binaries has shown that the poles can be absorbed into a renormalisation of the worldlines of the particles, so they should not appear in any physical coordinate invariant quantity. Thus dimensional regularization is a powerful regularization method in the PN context. In particular this regularization is free of the ambiguities plaguing the Hadamard regularization at the 3PN order \cite{DJSdim,BDE04,BDEI04} (see the contributions by Sch\"afer and Blanchet in this volume).

Although the two regularizations in SF and PN analyses have been carefully designed, it appears to be non trivial that they will yield results consistent up to a high level of approximation. Our cross cultural comparison of the redshift observable $u_1^T$ is a test of the equivalence of the SF and PN metrics \eqref{regSF1} and \eqref{gPNdr} and is, thus, a test of the two independent (and very different) regularization procedures in use.

\section{Circular orbits in the perturbed Schwarzschild geometry}
\label{overview}

Previously we described the truly coordinate and perturbative-gauge independent properties of $\Omega$ and the redshift observable $u_1^T$. In this section we use Schwarzschild coordinates, and we refer to ``gauge invariance'' as a property which holds within the restricted class of gauges for which \eqref{killing} is a helical Killing vector. In all other respects, the gauge choice is arbitrary. With this assumption, no generality is lost, and a great deal of simplicity is gained.

The effect of the gravitational self-force is most easily described as having $m_1$ move along a geodesic of the regularized metric \eqref{regSF}. We are interested in circular orbits and let $u\ua$ be the four-velocity of $m_1$.\footnote{Since we are interested in the motion of the small particle $m_1$, we remove the index $1$ from $u_1\ua$.} This differs from the four-velocity $\ubar\ua$ of a geodesic of the straight Schwarzschild geometry at the same radial coordinate $r$ by an amount of $\mathcal{O}(q)$. Recall that we are describing  perturbation analysis with $q\ll1$, therefore $h^\R\lab = \mathcal{O}(q)$, and all equations in this Section necessarily hold only through first order in $q$.

It is straightforward to determine the components of the geodesic equation for the metric \eqref{regSF} \cite{Det08}, and then to find the components of the four-velocity $u\ua$ of $\MU$ when it is in a circular orbit at Schwarzschild radius $r$. We reiterate that the four-velocity is to be normalized with respect to $\bar g\lab + h^\R\lab$ rather than $\bar g\lab$, and that $h^\R\lab$ is assumed to respect the symmetry of the helical Killing vector. In this case we have\footnote{In all of this section we shall set $G=c=1$.}
\begin{subequations}\label{uTuphi}
\begin{eqnarray}
  (u^t)^2 &=&
    \frac{{r}}{{r}-3{\m} }  \Big[ 1+\ubar^\alpha \ubar^\beta h^\R\lab
       - \frac{{r}}{2}\ubar^\alpha \ubar^\beta\partial_r  h^\R\lab \Big] \, ,
\label{uTinitial}
\\
  (u^\varphi)^2 &=&
      \frac{{r}-2{\m}}{{r}({r}-3{\m} )}
         \biggl[ \frac{{\m} (1+\ubar^\alpha \ubar^\beta h^\R\lab)}{{r}({r}-2{\m})}
     - \frac{1}{2} \ubar^\alpha \ubar^\beta\partial_r h^\R\lab \biggr]\,.
\label{uPeqn}
\end{eqnarray}\end{subequations}
A consequence of these relations is that the orbital frequency of $m_1$ in a circular orbit about a perturbed Schwarzschild black hole of mass $\m$ is, through first order in the perturbation, given by
\begin{equation}
  \Omega^2 = \biggl(\frac{u^\varphi}{u^t}\biggr)^2
     = \frac{{\m} }{{r}^3}
        - \frac{{r}-3{\m} }{2 {r}^2} \,\ubar\ua \ubar\ub \partial_r h^\R\lab\,.
\label{Omega2}
\end{equation}
The angular frequency $\Omega$ is a physical observable, and is independent of the gauge choice. However the perturbed Schwarzschild metric does not have spherical symmetry, and the radius of the orbit $r$ is not an observable and does depend upon the gauge choice. That is to say, an infinitesimal coordinate transformation of $\mathcal{O}(q)$ might change $\ubar\ua \ubar\ub \partial_r h^\R\lab$. But if it does, then it will also change the radius $r$ of the orbit in just such a way that $\Omega^2$ as determined from \eqref{Omega2} remains unchanged. Both $u^t\equiv u^T$ and  $u^\varphi \equiv \Omega \, u^T$ are gauge invariant as well.

Our principle interest is in the relationship between $\Omega$ and $u^T$, which we now establish directly using \eqref{uTinitial} and \eqref{Omega2}. To this end, following \cite{Det08}, we introduce the gauge invariant measure of the orbital radius
\begin{equation}
  R_\Omega \equiv \left( \frac{m_2}{\Omega^2} \right)^{1/3} \, ,
\label{ROmega}
\end{equation}
and readily establish a first order, gauge invariant, algebraic relationship between $u^T$ (to which $u^t$ evaluates in our gauge) and $R_\Omega$ (or equivalently $\Omega$), namely:
\begin{equation}
  (u^T)^2 =
    \left( 1 - \frac{3\m}{R_\Omega} \right)^{-1} \left( 1 + \ubar^\alpha \ubar^\beta h^\R\lab \right) .
\label{uT}
\end{equation}
See Paper I for a detailed derivation of this result.  
The lowest order term in $q$ on the right-hand-side is identical to what is obtained for a circular geodesic of the unperturbed Schwarzschild metric. Indeed, recall that the Schwarzschild part of $u^T$ is known exactly as $u^T_\mathrm{Schw} = \left( 1 - 3 m_2/R_\Omega \right)^{-1/2}$. Thus, if we write
\beq
u^T \equiv u^T_{\mathrm{Schw}}+q\, u^T_{\mathrm{SF}} + \mathcal{O}(q^2) \, ,
\eeq
the first order term in \eqref{uT} gives:
\beq
  q \, u^T_{\mathrm{SF}} = \frac12 \left(1-\frac{3m_2}{R_\Omega}\right)^{-1/2}\!\! \ubar\ua \ubar\ub h^\R\lab \,,
\label{SF}
\eeq
which is $\mathcal{O}(q)$, and contains the effect of the ``gravitational self-force'' on the relationship between $u^T$ and $\Omega$, even though it bears little resemblance to a force. The numerical SF approach henceforth focuses attention on the calculation of the combination $\ubar^\alpha \ubar^\beta h^\R\lab$. See \cite{Det08} and Paper I for more details on the implementation of the regularization of the perturbation, and on the numerical computation of the quantity $\ubar^\alpha \ubar^\beta h^\R\lab$.

\section{Overview of the 3PN calculation}
\label{3PN}

Our aim is to compute the 3PN regularized metric \eqref{gPNdr} by direct post-Newtonian iteration of the Einstein field equations in the case of singular point mass sources. In the dimensional regularization scheme, we look for the solution of the Einstein field equations in $d+1$ space-time dimensions. We treat the space dimension as an arbitrary complex number, $d\in\mathbb{C}$, and interpret any intermediate formula in the PN iteration of those equations by analytic continuation in $d$. Then we analytically continue $d$ down to the value of interest (namely 3), posing $d \equiv 3 + \varepsilon$. In most of the calculations we neglect terms of order $\varepsilon$ or higher, i.e. we retain the finite part and the eventual poles.

\subsection{Iterative PN computation of the metric}

Defining the gravitational field variable $h\ab\equiv\sqrt{-g}\, g\ab - \eta\ab$,\,\footnote{Here $g\ab$ is the contravariant metric, inverse of the covariant metric $g\lab$ of determinant $g = \text{det}(g\lab)$, and $\eta\ab=\mathrm{diag}(-1,1,1,1)$ represents an auxiliary Minkowski metric in Cartesian coordinates.} and adopting the harmonic coordinate condition $\partial_\beta h^{\alpha\beta}= 0$, we can write the ``relaxed'' Einstein field equations in the form of ordinary d'Alembert equations, namely
\begin{equation}\label{Dalembert}
\Box h\ab = \frac{16\pi G^{(d)}}{c^4} |g| \, T\ab + \Lambda\ab[h,\partial h, \partial^2 h]\,,
\end{equation}
where $\Box \equiv \eta^{\mu\nu}\partial_\mu\partial_\nu$ is the \textit{flat}-spacetime d'Alembertian operator in $d+1$ dimensions. The gravitational source term $\Lambda\ab$ in \eqref{Dalembert} is a functional of $h^{\mu\nu}$ and its first and second space-time derivatives; it depends explicitly on the dimension $d$. The matter stress-energy tensor $T\ab$ is composed of Dirac delta-functions in $d$ dimensions. Finally, the $d$-dimensional gravitational constant $G^{(d)}$ is related to the usual Newton constant $G$ by
\begin{equation}\label{Geps}
G^{(d)} = G\,\ell_0^\varepsilon\,,
\end{equation}
where $\ell_0$ denotes the characteristic length associated with dimensional regularization. We shall check that this length scale disappears from the final gauge-invariant 3-dimensional result.  

The 3PN metric is given in expanded form for general matter sources in terms of some ``elementary'' retarded potentials (sometimes called near-zone potentials) $V$, $V_i$, $K$, $\hat W_{ij}$, $\hat R_i$, $\hat X$, $\hat Z_{ij}$, $\hat Y_i$ and $\hat T$, which were introduced in Ref.~\cite{BFeom} for 3 dimensions and generalized to $d$ dimensions in Ref.~\cite{BDE04}. All these potentials have a finite non-zero post-Newtonian limit when $c \rightarrow +\infty$ and parameterize the successive PN approximations. Although this decomposition in terms of near-zone potentials is convenient, such potentials have no physical meaning by themselves. 

Let us first define the combination
\begin{equation}\label{calV}
    \mathcal{V} \equiv V -\frac{2}{c^2} \left(\frac{d-3}{d-2}\right) K + \frac{4 \hat X}{c^4} +\frac{16 \hat T}{c^6} \, .
\end{equation}
Then the 3PN metric components can be written in the rather compact form \cite{BDE04}
\begin{subequations}\label{metricexp}
    \begin{align}
        g^\text{PN}_{00} &= - e^{-2\mathcal{V}/c^2} \left( 1 - \frac{8 V_i V_i}{c^6} - \frac{32 \hat R_i V_i}{c^8} \right) + \mathcal{O}(c^{-10}) \, , \label{g00} \\
        g^\text{PN}_{0i} &= - e^{-\frac{(d-3)\mathcal{V}}{(d-2)c^2}} \left( \frac{4V_i}{c^3} \left[ 1 + \frac{1}{2} \left( \frac{d-1}{d-2} \frac{V}{c^2} \right)^2 \right] + \frac{8\hat R_i}{c^5} + \frac{16}{c^7} \left[ \hat Y_i + \frac{1}{2} \hat W_{ij} V_j \right] \right) + \mathcal{O}(c^{-9}) \, , \label{g0i} \\
        g^\text{PN}_{ij} &= e^{\frac{2\mathcal{V}}{(d-2)c^2}} \left( \delta_{ij} + \frac{4}{c^4} \hat W_{ij} + \frac{16}{c^6} \left[ \hat Z_{ij} - V_i V_j + \frac{1}{2(d-2)} \, \delta_{ij} V_k V_k \right] \right) + \mathcal{O}(c^{-8}) \, , \label{gij}
    \end{align}
\end{subequations}
where the exponentials are to be expanded to the order required for practical calculations.  The successive PN truncations of the field equations \eqref{Dalembert} give us the equations satisfied by all the above potentials up to 3PN order. We conveniently define from the components of the matter stress-energy tensor $T\ab$ the following density, current density, and stress density:
\begin{subequations}\label{sigma}
\begin{align}
    \sigma & \equiv \frac{2}{d-1} \frac{(d-2)T^{00}+T^{ii}}{c^2} \, , \\
    \sigma_i & \equiv \frac{T^{0i}}{c} \, , \\
    \sigma_{ij} & \equiv T^{ij} \, ,
\end{align}
\end{subequations}
where $T^{ii}\equiv\delta_{ij}T^{ij}$. As examples, the leading-order potentials in the metric obey
\begin{subequations}\label{potentialEq}
 \begin{align}
    \Box V = & - 4 \pi G^{(d)} \, \sigma \, , \label{dalV} \\
    \Box V_i = & - 4 \pi G^{(d)} \, \sigma_i \, , \label{dalVi} \\
    \Box\hat W_{ij} = & - 4 \pi G^{(d)} \biggl( \sigma_{ij} - \delta_{ij} \, \frac{\sigma_{kk}}{d-2} \biggr) - \frac{1}{2} \biggl( \frac{d-1}{d-2} \biggr) \partial_i V \partial_j V \, . \label{dalWij}
    \end{align}
\end{subequations}
All the potentials evidently include many PN corrections. The potentials $V$ and $V_i$ have a compact support (i.e. their source is localized on the isolated matter system) and will admit a finite limit when $\varepsilon\rightarrow 0$ without any pole. Most of the other potentials have, in addition to a compact-support part, a non-compact support contribution, such as that generated by the term $\propto \partial_i V \partial_j V$ in the source of $\hat W_{ij}$. These non-compact support pieces are the most delicate to compute, because they typically generate some poles $\propto 1/\varepsilon$ at the 3PN order. The d'Alembert equations satisfied by all higher-order PN potentials, whose sources are made of non-linear combinations of lower-order potentials, can be found in Paper I. Clearly, the higher the PN order of a potential, the more complicated is its source, but it requires computations at a lower relative order.

Many of the latter potentials have already been computed for compact binary systems, and we have extensively used these results from \cite{BFeom,BDE04}. Notably, all the compact-support potentials such as $V$ and $V_i$, and all the compact-support parts of other potentials, have been computed for any field point $\mathbf{x}$, and then at the source point $\mathbf{y}_1$ following the regularization. However, the most difficult non-compact support potentials such as $\hat X$ and $\hat T$ could not be computed at any field point $\mathbf{x}$, and were regularized directly on the particle's world-line. Since for the equations of motion we needed only the \textit{gradients} of these potentials, only the gradients were regularized on the particle, yielding the results for ${(\partial_i\hat X)}_1$ and ${(\partial_i\hat T)}_1$ needed in the equations of motion. However the 3PN metric requires the values of the potentials themselves regularized on the particles, i.e. ${(\hat X)}_1$ and ${(\hat T)}_1$. For the present work we therefore computed, using the tools developed in \cite{BFeom,BDE04}, the difficult non-linear potentials ${(\hat X)}_1$ and ${(\hat T)}_1$, and especially the non-compact support parts therein. Unfortunately, the potential $\hat X$ is always the most tricky to compute, because its source involves the cubically-non-linear and non-compact-support term $\hat W_{ij}\, \partial_{ij}V$, and it has to be evaluated at relative 1PN order.

In this calculation we also met a new difficulty with respect to the computation of the 3PN equations of motion. Indeed, we found that the potential $\hat X$ is divergent because of the bound of the Poisson-like integral at \textit{infinity}. Thus the potential $\hat X$ develops an IR divergence, in addition to the UV divergence due to the singular nature of the source and which is cured by dimensional regularization. The IR divergence is a particular case of the well-known divergence of Poisson integrals in the PN expansion for general (regular) sources, linked to the fact that the PN expansion is a singular perturbation expansion, with coefficients typically blowing up at spatial infinity. The IR divergence is discussed in Paper I, where we show how to resolve it by means of a finite part prescription.

The 3PN metric \eqref{metricexp} is valid for a general isolated matter system, and we apply it to the case of a system of $N$ point-particles with ``Schwarzschild'' masses $m_a$ and without spins (here $a=1,\cdots,N$). In this case we have
\begin{subequations}
\begin{align}
\sigma (\mathbf{x} , t) &= \sum_a \tilde\mu_a  \, \delta^{(d)} [\mathbf{x} - \mathbf{y}_a (t)]\,,\\\sigma_{i} (\mathbf{x} , t) &= \sum_a \mu_a \,v_a^i\, \delta^{(d)} [\mathbf{x} - \mathbf{y}_a (t)]\,,\\\sigma_{ij} (\mathbf{x} , t) &= \sum_a \mu_a \,v_a^i\,v_a^j \,\delta^{(d)} [\mathbf{x} - \mathbf{y}_a (t)]\,,
\end{align}
\end{subequations}
where $\delta^{(d)}$ denotes the Dirac density distribution in $d$ spatial dimensions, such that $\int \ud^d \mathbf{x} \, \delta^{(d)}(\mathbf{x}) = 1$. We defined the effective masses of the particles by
\begin{subequations}
\begin{align}
\mu_a (t) &= \frac{m_a}{\sqrt{(gg_{\alpha \beta}) (\mathbf{y}_a, t) \, v_a^{\alpha} \, v_a^{\beta}/c^2}} \,,\\
\tilde\mu_a (t) &= \frac{2}{d-1} \left[ d - 2 + \frac{\mathbf{v}_a^2}{c^2} \right] \mu_a(t)\,.
\end{align}
\end{subequations}

\subsection{The example of the zeroth-order iteration}

For illustration purposes, let us consider the simple case of the dimensional regularization of the Newtonian potential generated by two point particles of masses $m_1$ and $m_2$. The PN metric \eqref{metricexp} reduces to $g^\text{PN}_{00} = -1 + 2V/c^2 + \mathcal{O}(c^{-4})$, where $V$ is solution of the Poisson equation $\Delta V = -4 \pi G^{(d)} \sigma + \mathcal{O}(c^{-2})$, with source $\sigma(\mathbf{x}) = \frac{2(d-2)}{d-1} \bigl[ m_1 \delta^{(d)}(\mathbf{x} - \mathbf{y}_1) + m_2 \delta^{(d)}(\mathbf{x} - \mathbf{y}_2) \bigr] + \mathcal{O}(c^{-2})$. Solving this Poisson equation in $d$ space dimensions yields 
\begin{equation}\label{V_dim_d}
	V(\mathbf{x}) = \frac{2(d-2)}{d-1} \, k \, \biggl( \frac{G^{(d)} m_1}{\vert \mathbf{x} - \mathbf{y}_1 \vert^{d-2}} + \frac{G^{(d)} m_2}{\vert \mathbf{x} - \mathbf{y}_2 \vert^{d-2}} \biggr) + \mathcal{O}(c^{-2})\, ,
\end{equation}
where $k \equiv \Gamma(\frac{d-2}{2})/\pi^\frac{d-2}{2}$ tends to $1$ when $\varepsilon \rightarrow 0$ ($\Gamma$ is the usual Eulerian function). When $d=3$, the Newtonian potential \eqref{V_dim_d} is not defined in the limit $\mathbf{x} \rightarrow \mathbf{y}_1$ because of the divergent self-field of particle $1$. By contrast, thanks to the analytic continuation in the space dimension, it is always possible to choose $\Re(d) < 2$ such that the $d$-dimensional potential \eqref{V_dim_d} has a well-defined limit when $\mathbf{x} \rightarrow \mathbf{y}_1$, namely ${(V)}_1 \equiv V(\mathbf{y}_1)$ given by
\begin{equation}
	{(V)}_1 = \frac{2(d-2)}{d-1} \, k \, \frac{G^{(d)} m_2}{r_{12}^{d-2}} + \mathcal{O}(c^{-2}) \, ,
\end{equation}
where we have posed $r_{12}\equiv \vert \mathbf{y}_1 - \mathbf{y}_2 \vert$. Relying on the unicity of the analytic continuation, we obtain the unique 3-dimensional result
\begin{equation}\label{V1}
	{(V)}_1 = \frac{G m_2}{r_{12}} + \mathcal{O}(\varepsilon) + \mathcal{O}(c^{-2})\, .
\end{equation}

This procedure is clear up to a high order, but let us mention a subtle point in the calculation of Paper I, namely that we had to systematically re-introduce the correction terms $\mathcal{O}(\varepsilon)$ in the \textit{Newtonian} part of the metric and other quantities. Indeed, in various operations such as replacing $r_{12}$ in Eq.~\eqref{V1} by its 3PN expression in terms of the orbital frequency $\Omega$, the corrections $\mathcal{O}(\varepsilon)$ will be multiplied by some poles at 3PN order, and they will therefore contribute \textit{in fine} to the finite part at 3PN order. Such corrections are necessary only in the Newtonian results (since the poles arise only at 3PN order). For instance, the 3-dimensional Newtonian potential \eqref{V1} is to be replaced by its $d$-dimensional version valid up to terms $\mathcal{O}(\varepsilon^2)$, namely 
\begin{equation}\label{UNdexp}
    {(V)}_1 = \frac{G m_2}{r_{12}} \left\{ 1 + \varepsilon \left[ \frac{1}{2} - \ln{\left( \frac{r_{12} \, p}{\ell_0} \right)} \right] + \mathcal{O}(\varepsilon^2)\right\} + \mathcal{O}(c^{-2}) \,,
\end{equation}
in which $p \equiv \sqrt{4 \pi} \,e^{C/2}$ with $C = 0.5772\cdots$ being the Euler--Mascheroni constant.

This Newtonian calculation is generalized at higher orders in the iterative PN process described in the previous Section. The result is the 3PN regularized metric \eqref{g1dr} which is given in explicit form by Eqs.~(4.2) of Paper I.

\section{Logarithmic terms at 4PN and 5PN orders}
\label{logs}

We now discuss the logarithmic contributions in the near-zone metric of a generic isolated post-Newtonian source, and then of a compact binary system. Our motivation is that knowing analytically determined \textit{logarithmic} terms in the PN expansion is crucial for efficiently extracting from the SF data the numerical values of higher order PN coefficients.  This will be further discussed in Sec. \ref{secVI}.

The occurence of logarithmic terms in the PN expansion has been investigated in many previous works \cite{Anderson,K80a,K80b,AKKM82,FS83,F83,BD86,BD88,B93}. Notably Anderson \textit{et al.} \cite{AKKM82} found that the dominant logarithm arises at the 4PN order, and Blanchet \& Damour \cite{BD88,B93} showed that this logarithm is associated with gravitational wave tails modifying the usual 2.5PN radiation-reaction damping at the 4PN order. Furthermore the general structure of the PN expansion is known \cite{BD86}: it is of the type $\sum (v/c)^k[\ln(v/c)]^q$, where $k$ and $q$ are positive integers, involving only powers of logarithms; more exotic terms such as $[\ln(\ln(v/c))]^q$ cannot arise. 

Following Paper II, we shall determine the leading 4PN logarithm and the next-to-leading 5PN logarithm in the conservative part of the dynamics of a compact binary system. The computation of such logarithmic contributions relies only very weakly on a regularization scheme. We shall thus work in $3$ space dimensions; from now on we set $\varepsilon = 0$. 

\subsection{Physical origin of logarithmic terms}
\label{logphys}

Because of the non-linearity of the field equations, the gravitational field at coordinate time $t$ is in general not a function of the state of motion of the source at retarded time $t - r /c$, where $r= \vert \mathbf{x} \vert$ is the distance to the center of the source, but depends on the entire past ``history'' of the source. This means that the near-zone metric depends on the source at all times before the current time $t$, say $t'\leqslant t$ (indeed in the near-zone one expands all retardations, i.e. $r/c\to 0$). This ``hereditary'' effect starts at 4PN order in the near-zone, and originates from gravitational-wave tails, namely the scattering of gravitational radiation by the background curvature of the spacetime generated by the mass $M$ of the source. In a certain gauge where the hereditary terms are collected into the $00$ component of the metric, we have
\begin{equation}\label{g00_tail}
	\delta g_{00}^\text{tail}(\mathbf{x},t) = - \frac{8G^2M}{5c^{10}} \,x^a x^b \int_{-\infty}^t \ud t' \,M_{ab}^{(7)}(t') \ln{\left( \frac{c(t - t')}{2r} \right)} \,,
\end{equation}
where $M$ is the ADM mass of the source, and $M_{ab}^{(n)}$ is the $n$-th time derivative of its quadrupole moment. This tail-induced contribution is of 4PN order.

The occurence of the tail effect in the near-zone metric implies that the usual 2.5PN radiation-reaction force density in the matter source is corrected at the relative 1.5PN order as
\begin{equation}\label{F_rad_reac}
	F^i_\text{rad}(\mathbf{x},t) = - \frac{2G}{5c^5} \,\rho\,x^a \left[M_{ia}^{(5)}(t) + \frac{4G M}{c^3} \int_{-\infty}^t \ud t' \,M_{ia}^{(7)}(t') \ln{\left( \frac{c(t - t')}{2 \lambda} \right)} \right]\, ,
\end{equation}
where $\rho$ is the Newtonian mass density in the source, and $\lambda$ is the typical wavelength of the radiation. The leading term in \eqref{F_rad_reac} is the standard Burke-Thorne \cite{Bu71,BuTh70} radiation-reaction potential at the 2.5PN order, responsible for the leading radiation effect. The hereditary correction was obtained in \cite{BD88,B93}, and shown to be consistent with wave tails propagating at large distances from the source \cite{BD92}.

The radiation-reaction force \eqref{F_rad_reac} deserves its name because it is not invariant under a time reversal, and therefore gives rise to dissipative effects. A good way to see this is to change the condition  of retarded potentials to advanced potentials, i.e. to formally change $c$ into $-c$. The first term in \eqref{F_rad_reac} is clearly non invariant because it comes with an odd number of powers of $1/c$. The second term is also non invariant, despite the fact that it comes with an even power of $1/c$ in front (i.e. $1/c^8$ in \eqref{F_rad_reac}, corresponding to 4PN), because it is composed of an integral extending over the past rather than a time-symmetric integral. 

However, thanks to the even power of $1/c$ carried by the hereditary integral \eqref{g00_tail}, this means that there exists a \textit{conservative} piece associated with it. Recall that in our calculation we neglect the dissipative radiation-reaction effects and are interested only in the conservative part of the dynamics;  we have implemented this restriction by assuming the existence of the helical Killing vector \eqref{killing}, depending on the orbital frequency $\Omega$ of the circular motion. The presence of this scale $\Omega$, imposed by the helical Killing symmetry, permits immediately to identify the conservative piece associated with the tail term in \eqref{g00_tail}, through the decomposition
\begin{equation}\label{log}
	\ln{\left(\frac{c(t - t')}{2 r}\right)} = \ln{\left(\frac{c(t - t')}{2 \lambda}\right)} - \ln{\left(\frac{r}{\lambda}\right)}\,,
\end{equation}
where now the characteristic length scale $\lambda$ is defined by $\lambda \equiv 2 \pi c/\Omega$. The second term in \eqref{log}, when inserted into the tail integral in \eqref{g00_tail}, can be integrated out and gives rise to the conservative 4PN piece
\begin{equation}\label{g004PN}
	\delta g_{00}^\text{cons}(\mathbf{x},t) = \frac{8G^2 M}{5c^{10}} \,x^a x^b \,M_{ab}^{(6)}(t)\,\ln{\left(\frac{r}{\lambda}\right)} \,.
\end{equation}
This contribution to the metric has to be included in our study of the conservative part of the dynamics, while the purely dissipative piece in \eqref{g00_tail} enters the radiation reaction \eqref{F_rad_reac}, and is excluded by our assumption of existence of the helical Killing vector. The term \eqref{g004PN} is precisely the conservative 4PN logarithmic contribution in the near-zone metric that we want to compute, as well as its 5PN correction.

Notice that it is possible to find a gauge where all the logarithms are gathered in the $00$ component of the metric, as we did in Eq.~\eqref{g004PN}, only at 4PN order. When looking to subdominant logarithmic terms at 5PN order, we are obliged to include some vectorial $0i$ and tensorial $ij$ components of the metric. But still it will be possible, and extremely convenient, to define a gauge in which the logarithms are ``maximally'' transferred to the $00$ and $0i$ components (while the remaining contribution in the $ij$ components is ``minimal''). Using such a gauge
saves a lot of calculations when performing the PN iteration; this was the strategy adopted in Paper II to compute the higher order 5PN logarithms beyond Eq.~\eqref{g004PN}.

\subsection{Expression of the near-zone metric}
\label{nzres}

To compute these 4PN and 5PN conservative logarithmic
contributions, we make use of the multipolar post-Minkowskian wave-generation
formalism \cite{BD86,BD88,BD92,B93}. We first identify these logarithmic
contributions in the exterior of a generic isolated post-Newtonian source. We
then deduce the metric inside the matter source by a matching performed in the
exterior part of the near-zone. We finally get the 4PN and 5PN logarithmic
contributions in the near-zone metric, valid in a specific gauge
defined in Paper II, as\footnote{We use shorthands such as $x^{ab} = x^a x^b$; $\hat x^{abc} = x^{abc} - \frac{1}{5} (\delta^{ab} x^c + \delta^{ac} x^b + \delta^{bc} x^a) r^2$ denotes the symmetric and trace-free part of $x^{abc}$; $\varepsilon_{abc}$ is the Levi-Civita antisymmetric symbol.}
\begin{subequations}\label{glog}
\begin{align}
\delta{g}_{00} &= \frac{G^2M}{c^{10}}\left[\frac{8}{5}\left(1-\frac{2U}{c^2}\right)x^{ab}M^{(6)}_{ab}+\frac{4}{35c^2}r^2x^{ab}M^{(8)}_{ab}-\frac{8}{189c^2}x^{abc}M^{(8)}_{abc}\right]\ln\left(\frac{r}{\lambda}\right) \nonumber \\ &- \frac{8}{5}\frac{G^3M}{c^{12}}x^{a}M^{(6)}_{ab}\int\frac{\ud^3x'}{\vert\mathbf{x}-\mathbf{x}'\vert}\,\rho'\,x'^{b}\ln\left(\frac{r'}{\lambda}\right)+ \mathcal{O}\left(\frac{1}{c^{14}}\right)\,,\label{deltag00}\\
\delta{g}_{0i} &= \frac{G^2M}{c^{11}}\left[\frac{16}{21}\hat{x}^{iab}M^{(7)}_{ab}-\frac{64}{45}\varepsilon_{iab}x^{ac}S^{(6)}_{bc}\right]\ln\left(\frac{r}{\lambda}\right) + \mathcal{O}\left(\frac{1}{c^{13}}\right)\,,\\
\delta{g}_{ij} &= \frac{G^2M}{c^{10}}\left[\frac{8}{5}x^{ab}M^{(6)}_{ab}\delta_{ij}\right]\ln\left(\frac{r}{\lambda}\right) + \mathcal{O}\left(\frac{1}{c^{12}}\right)\,,
\end{align}\end{subequations}
where $M_{ab}$, $M_{abc}$ and $S_{ab}$ denote the mass quadrupole, mass octupole and current quadrupole moments of the source respectively, and where 
\begin{align}
U(\mathbf{x},t) = G \int\frac{\ud^3x'}{\vert\mathbf{x}-\mathbf{x}'\vert}\,\rho(\mathbf{x}',t)
\end{align}
is the Newtonian potential, sourced by the Newtonian mass density $\rho$ of the source. We recall that $\lambda = 2 \pi c/\Omega$, where $\Omega$ is the scale entering the Killing vector \eqref{killing}. Notice in particular the 5PN contribution in $\delta{g}_{00}$ which involves the Poisson integral of a logarithmically modified source density, and which will contribute \textit{in fine} to the 5PN logarithms in the case of binary systems.

We can then apply the result \eqref{glog} to the case of a compact binary system moving on a circular orbit. In this case, $\Omega$ is the orbital frequency of the motion. We compute the metric at the location of one of the particles and obtain the result
\begin{subequations}
	\begin{align}
		{(\delta{g}_{00})}_1 &= \frac{G^2M}{c^{10}}\left[\frac{8}{5}\left(1-\frac{2U_1}{c^2}\right)y_1^{ab}M^{(6)}_{ab}-\frac{8}{5c^2}U_1y_1^{a}y_2^{b}M^{(6)}_{ab}\right.\nonumber\\&\left.\qquad\quad+\frac{4}{35c^2}y_1^2y_1^{ab}M^{(8)}_{ab}-\frac{8}{189c^2}y_1^{abc}M^{(8)}_{abc}\right]\ln\left(\frac{r_{12}}{\lambda}\right)+ \mathcal{O}\left(\frac{1}{c^{14}}\right),\\
		{(\delta{g}_{0i})}_1 &= \frac{G^2M}{c^{11}}\left[\frac{16}{21}\hat{y}_1^{iab}M^{(7)}_{ab}-\frac{64}{45}\varepsilon_{iab}\,y_1^{ac}S^{(6)}_{bc}\right]\ln\left(\frac{r_{12}}{\lambda}\right)+ \mathcal{O}\left(\frac{1}{c^{13}}\right),\\
		{(\delta{g}_{ij})}_1 &= \frac{G^2M}{c^{10}}\left[\frac{8}{5}y_1^{ab}M^{(6)}_{ab}\delta_{ij}\right]\ln\left(\frac{r_{12}}{\lambda}\right)+ \mathcal{O}\left(\frac{1}{c^{12}}\right),
	\end{align}
\end{subequations}
where $U_1 = G m_2 / r_{12}$ is the Newtonian potential felt by particle 1. Finally, the last step is to replace the multipole moments $M_{ab}$, $M_{abc}$ and $S_{ab}$ by the relevant PN expressions valid for circular-orbit compact binaries; we need also the ADM mass $M$, which reduces to $m=m_1+m_2$ in first approximation. Note that the mass quadrupole moment $M_{ab}$ (and also the ADM mass $M$) must crucially include a 1PN contribution. Again, we emphasize that for the 4PN and 5PN logarithms we do not need the full apparatus of dimensional regularization, in contrast to the fully fledged 3PN calculation sketched in Sec. \ref{3PN}. 

\section{Post-Newtonian results for the redshift observable}
\label{PN}

To compute the gauge invariant quantity $u^T$ (associated with the particle 1 for helical symmetry, circular orbits), we adopt its coordinate form as given by \eqref{ut_defX}, namely
\begin{equation}\label{utPN}
    u^t = \biggl( - {(g\lab)}_1 \frac{v_1\ua v_1\ub}{c^2} \biggr)^{-1/2} \, ,
\end{equation}
and plug into it the 3PN regularized metric \eqref{g1dr}, explicitly computed from the 3PN near-zone expression \eqref{metricexp} reduced to binary point masses, and including the 4PN and 5PN logarithmic corrections computed in Sec.~\ref{logs}. To begin with, this yields the expression of $u^t$ for an arbitrary mass ratio $q=m_1/m_2$, and for a generic non-circular orbit in a general reference frame. We then choose the frame of the center of mass, which is consistently defined by the nullity of the center-of-mass integral of the motion, deduced from the equations of motion. Restricting ourselves to exactly circular orbits (consistently with the helical Killing symmetry we neglect radiation-reaction effects), the result is expressed by means of the convenient dimensionless gauge invariant PN parameter
\begin{equation}\label{x}
    x \equiv \left( \frac{G\,m\,\Omega}{c^3} \right)^{2/3} \, ,
\end{equation}
which is directly related to the orbital frequency $\Omega$ of the circular orbit, and depends on the total mass $m = m_1 + m_2$ of the binary.

We discover most satisfactorily that all the poles $\propto 1/\varepsilon$ (as well as the associated constant $\ell_0$) cancel out in the final expression for $u^t$. Our final result for a 3PN (plus 4PN and 5PN logarithmic terms), gauge invariant, algebraic relationship between $u^T$ (to which $u^t$ now evaluates) and $x$ (or equivalently $\Omega$), is\footnote{The Landau $o$ symbol for remainders takes its standard meaning.}
\begin{align}\label{uT_PN}
    u^T(x) &= 1 + \left( \frac{3}{4} + \frac{3}{4} \Delta - \frac{\nu}{2} \right) x + \left( \frac{27}{16} + \frac{27}{16} \Delta - \frac{5}{2} \nu - \frac{5}{8} \Delta \, \nu + \frac{\nu^2}{24} \right) x^2 \nonumber \\ &+ \left( \frac{135}{32} + \frac{135}{32} \Delta - \frac{37}{4} \nu - \frac{67}{16} \Delta \, \nu + \frac{115}{32} \nu^2 + \frac{5}{32} \Delta \, \nu^2 + \frac{\nu^3}{48} \right) x^3 \nonumber \\ &+ \left( \frac{2835}{256} + \frac{2835}{256} \Delta - \left[ \frac{2183}{48} - \frac{41}{64} \pi^2 \right] \nu - \left[ \frac{12199}{384} - \frac{41}{64} \pi^2 \right] \Delta \, \nu \right. \nonumber \\ &\left. + \left[ \frac{17201}{576} - \frac{41}{192} \pi^2 \right] \nu^2 + \frac{795}{128} \Delta \, \nu^2 - \frac{2827}{864} \nu^3 + \frac{25}{1728} \Delta \, \nu^3 + \frac{35}{10368} \nu^4 \right) x^4 \nonumber \\ &+ \left( A_4(\nu) + \left[ - \frac{32}{5} - \frac{32}{5} \Delta + \frac{64}{15} \nu \right] \nu \, \ln x \right) x^5 \nonumber \\ &+ \left( A_5(\nu) + \left[ \frac{478}{105} + \frac{478}{105} \Delta + \frac{1684}{21} \nu + \frac{4388}{105} \Delta \nu - \frac{3664}{105} \nu^2 \right] \nu \, \ln x \right) x^6 \nonumber \\ &+ o(x^6) \, .
\end{align}
We introduced the notation $\Delta \equiv (m_2-m_1)/m = \sqrt{1 - 4 \nu}$, where $\nu = m_1 m_2 / m^2$ is the symmetric mass ratio. While it has been shown in \cite{Det08} (see also Sec.~\ref{method} above) that $u^T$ is gauge invariant at any PN order in the extreme mass ratio limit $\nu\ll 1$, here we find that it is also gauge invariant for \textit{any} mass ratio up to 3PN order (even up to 5PN order for the logarithmic terms). This result is expected from \eqref{ut_def}, according to which $u^T$ is a scalar under our hypothesis of helical symmetry. Being proportional to the symmetric mass ratio $\nu$, the 4PN and 5PN logarithmic contributions vanish in the test-mass limit --- this is clear given that the Schwarzschild result for $u^T(\Omega)$ does not involve any logarithm. Notice that the functions $A_4(\nu)$ and $A_5(\nu)$ entering the expression of the non-logarithmic contribution to $u^T(\Omega)$ at the 4PN and 5PN orders are unknown, and would be very difficult to compute within standard PN theory. However we know that they are polynomials in $\nu$, with leading-order coefficient given by the Schwarzschildean result [see Eqs.~\eqref{A4A5}].

We now investigate the small mass ratio regime $q\ll 1$, for comparison purposes with the perturbative SF calculation. We introduce a convenient PN parameter appropriate to the small mass limit of particle 1:
\begin{equation}\label{y}
    y \equiv \left( \frac{G\,m_2\,\Omega}{c^3} \right)^{2/3} \, ,
\end{equation}
which is related to the usual PN parameter $x$ by $x = y(1+q)^{2/3}$, and to the gauge-invariant measure \eqref{ROmega} of the orbital radius by $y = G m_2 / (R_\Omega c^2)$. We also use the expression of the symmetric mass ratio $\nu$ in terms of the (asymmetric) mass ratio $q = m_1/m_2$, namely $\nu = q / (1+q)^2$. Our complete redshift observable, expanded through post-self-force order, is of the type
\begin{equation}\label{utexp}
    u^T = u^T_\mathrm{Schw} + q \, u^T_\mathrm{SF}  + q^2 \, u^T_\mathrm{PSF} + \mathcal{O}(q^3) \,,
\end{equation}
where the Schwarzschild result is known in closed form as $u^T_\mathrm{Schw} = \left( 1 - 3 y \right)^{-1/2}$. By expanding the PN result \eqref{uT_PN} in powers of $q$, we find that the self-force contribution reads
\begin{align}\label{utSF}
u^T_\mathrm{SF}(y) &= - y - 2 y^2 - 5 y^3 + \left(
- \frac{121}{3} + \frac{41}{32} \pi^2 \right) y^4 \nonumber\\&+ \left(
\alpha_4 - \frac{64}{5}\ln y\right) y^5 + \left(
\alpha_5 + \frac{956}{105}\ln y\right) y^6 + o(y^6)\,.
\end{align}
The coefficients $\alpha_4$ and $\alpha_5$ are pure numbers which parametrize the small mass ratio expansions of the functions $A_4$ and $A_5$ through
\begin{subequations}\label{A4A5}
	\begin{align}
		A_4 &= \frac{15309}{256} + \left( \alpha_4 - \frac{25515}{128} \right) q + \mathcal{O}(q^2) \, , \\
		A_5 &= \frac{168399}{1024} + \left( \alpha_5 - \frac{168399}{256} \right) q + \mathcal{O}(q^2) \, .
	\end{align}
\end{subequations}

We also give the result for the combination $\ubar^\alpha \ubar^\beta h^\R\lab$ related to $u^T_\mathrm{SF}$ by Eq.~\eqref{SF}, since this is the quantity primarily used in the numerical SF calculation:
\begin{align}\label{uuhSF}
\ubar^\alpha \ubar^\beta \hat{h}^\R\lab &= - 2 y - y^2 - \frac{7}{4} y^3 + \left(
- \frac{1387}{24} + \frac{41}{16} \pi^2 \right) y^4 \nonumber\\&+ \left(
a_4 - \frac{128}{5}\ln y\right) y^5 + \left(
a_5 + \frac{5944}{105}\ln y\right) y^6 + o(y^6)\,.
\end{align}
We have conveniently rescaled the first-order perturbation $h^\R\lab$ by the mass ratio $q$, denoting $\hat{h}^\R\lab \equiv h^\R\lab / q$. Here $a_4$ and $a_5$ denote some unknown pure numbers related to $\alpha_4$ and $\alpha_5$ by
\begin{subequations}
	\begin{align}
		a_4 &= 2 \alpha_4 + \frac{9301}{64} - \frac{123}{32} \pi^2 \, , \\
		a_5 &= 2 \alpha_5 - 3 \alpha_4 + \frac{17097}{128} - \frac{369}{128} \pi^2 \, .
	\end{align}
\end{subequations}
The expansions \eqref{utSF}--\eqref{uuhSF} were determined up to 2PN order $\propto y^3$ in \cite{Det08}, based on the Hadamard-regularized 2PN metric given in \cite{BFP98}. The result at 3PN order $\propto y^4$ was obtained in Paper I using the powerful dimensional regularization scheme. By comparing the expansion \eqref{utSF} with our accurate numerical SF data for $u^T_\text{SF}(\Omega)$, we shall be able to measure the coefficients $\alpha_4$ and $\alpha_5$ (or $a_4$ and $a_5$) with at least 8 significant digits for the 4PN coefficient, and 5 significant digits for the 5PN coefficient. These results, as well as the estimation of even higher-order PN coefficients, will be detailed in the next Section.

Similarly, from the PN result \eqref{uT_PN} valid for any mass ratio $q$, we get the post-self-force contribution as
\begin{align}\label{utpSF}
u^T_\mathrm{PSF}(y) &= y + 3 y^2 + \frac{97}{8} y^3 + \left(
        \frac{725}{12} - \frac{41}{64} \pi^2 \right) y^4 \nonumber\\&+ \epsilon_4 \,y^5 + \left(\epsilon_5 + \frac{4588}{35}\ln y\right) y^6 + o(y^6) \, ,
\end{align}
which could in principle be compared to a future post-self-force calculation making use of second-order black hole perturbation theory. Note that there is no logarithm at 4PN order in the post-self-force term; the next 4PN logarithm would arise at cubic order $q^3$, i.e. at the post-post-SF level. The coefficients $\epsilon_4$ and $\epsilon_5$ in \eqref{utpSF} are unknown, and unfortunately they are expected to be extremely difficult to obtain, not only analytically in the standard PN theory, but also numerically as they require a second-order perturbative SF scheme.

\section{Numerical evaluation of post-Newtonian coefficients}\label{secVI}

In the self-force limit, the SF effect $\uT_\SF$ on the redshift observable $\uT$ is related via (\ref{uT}) to the regularized metric perturbation $\hhR$ at the location of the particle through
\begin{equation}
 u^T_\SF =
    \frac12 (1 - 3y)^{-1/2} \, \ubar^\alpha \ubar^\beta \hat{h}^\R\lab \, ,
\label{uTeqn}
\end{equation}
where $\ubar^\alpha$ is the background four-velocity of the particle. Recall that here $\hhR$ stands for the perturbation \textit{per unit mass ratio}, that is $\hR/q$. In SF analysis, the combination $\uuhh$ arises more naturally than $\uT_\SF$; this is the quantity we shall be interested in fitting in this Section. However our final results in Table \ref{bestfit} will include the corresponding values of the coefficients for the redshift observable $\uT_\SF$. We refer to Sec. II of Paper I for a discussion of the computation of the regularized metric perturbation $\hhR$, and the invariant properties of the combination $\uuhh$ with respect to the choice of perturbative gauge. In this Section we often use $r = 1/y$, a gauge invariant measure of the orbital radius scaled by the black hole mass $m_2$ [see Eqs.~\eqref{ROmega} and \eqref{y}].

Our earlier numerical work in \cite{Det08,BDLW10a,BDLW10b} provided values of the function $\uuhh(r)$ which cover a range in $r$ from $4$ to $750$. Following a procedure described in \cite{DMW03}, we have used Monte Carlo analysis to estimate the accuracy of our values for $\uuhh$. As was reported in Paper I, this gives us confidence in these base numbers to better than one part in $10^{13}$. We denote a standard error $\sigma$ representing the numerical error in $\uuhh$ by
\begin{equation}
  \sigma \simeq  |\uuhh| \times {\rm E}\times10^{-13},
\label{asymp}
\end{equation}
where $\rm E \simeq 1$ is being used as a placeholder to identify our estimate of the errors in our numerical results.

\subsection{Overview}
\label{secVIA}

A common task in physics is creating a functional model for a set of data. In our problem we have a set of $N$ data points $f_i$ and associated uncertainties $\sigma_i$, with each pair evaluated at an abscissa $r_i$. We wish to represent this data as some model function $f(r)$ which consists of a linear sum of $M$ basis functions $F_j(r)$ such that
\begin{equation}
  f(r) = \sum_{j=1}^M c_j F_j(r) \, .
\end{equation}
The numerical goal is to determine the $M$ coefficients $c_j$ which yield the best fit in a least squares sense over the range of data. That is, the $c_j$ are to be chosen such that
\begin{equation}
  \chsq \equiv \sum_{i=1}^N
       \left[\frac{f_i - \sum_{j=1}^M c_j F_j(r_i)}{\sigma_i}\right]^2
\label{chi2def}
\end{equation}
is a minimum under small changes in the $c_j$. For our application we choose the basis functions $F_j(r)$ to be a set of terms which are typical in PN expansions, such as $r^{-1}$, $r^{-2}$, \ldots, and also terms such as $r^{-5}\ln(r)$. We recognize that a solution to this extremum problem is not guaranteed to provide an accurate representation of the data $(r_i, f_i,\sigma_i)$. The quality of the numerical fit is measured by $\chsq$ as defined in Eq.~(\ref{chi2def}). If the model of the data is a good one, then the $\chi^2$ statistic itself has an expectation value of the number of degrees of freedom in the problem, $N-M$, with an uncertainty (standard deviation) of $\sqrt{2(N-M)}$.

Our numerical work leans heavily upon Ref.~\cite{PrTeVeFl} for solving the extremum problem for Eq.~(\ref{chi2def}). The numerical evaluation of the fitting coefficient $c_j$ includes a determination of its uncertainty $\Sigma_j$ which depends upon i) the actual values of $r_i$ in use, ii) all of the $\sigma_i$, and iii) the set of basis functions $F_j(r)$. 
In fact, the estimates of the $\Sigma_j$ do not depend at all on the data (or residuals) being fitted. As a consequence the estimates of the $\Sigma_j$ are only valid if the data are well represented by the set of basis functions. For emphasis: the $\Sigma_j$ depend upon $F_j(r_i)$ and upon $\sigma_i$ but \textit{are completely independent} of the $f_i$. Only if the fit is considered to be good, could the $\Sigma_j$ give any kind of realistic estimate for the uncertainty in the coefficients $c_j$. If the fit is not of high quality (unacceptable $\chsq$), then the $\Sigma_j$ bear no useful information \cite{PrTeVeFl}. We will come back to this point in the discussion below.

We also should remark that the task of determining coefficients in the $1/r$ characterization of our numerical data is almost incompatible with the task of determining an asymptotic expansion of $\uuhh$ from an analytic analysis. Analytically, the strict $r\rightarrow+\infty$ limit is always technically possible, whereas numerically, not only is that limit {\it never} attainable, but we must always contend with function evaluations at just a finite number of discrete points, obtained within a finite range of the independent variable, and computed with finite numerical precision.  Nevertheless, this is what we have done below.  %intend to do.

The numerical problem is even more constrained than we have just indicated. At large $r$, even though the data may still be computable there, the higher order terms for which we are interested in evaluating PN coefficients rapidly descend below the error level of our numerical data. This is clearly evident in Fig.~\ref{bestfig} below. For small $r$, the introduction of so many PN coefficients is necessary that it becomes extremely difficult to characterize our numerical data accurately. Thus, in practice, we find ourselves actually working with less than the full range of our available data. At large $r$ we could effectively drop points because they contribute so little to any fit we consider. At the other extreme, the advantage of adding more points in going to smaller $r$ is rapidly outweighed by the increased uncertainty in every fitted coefficient. This results from the need to add more basis functions in an attempt to fit the data at small $r$. Further details will become evident in Sec.~\ref{secVID} below.

\subsection{Framework for evaluating PN coefficients numerically}
\label{secVIB}

In a generic fashion we describe an expansion of $\ubar^\alpha \ubar^\beta \hat{h}^\R\lab$ in terms of PN coefficients $a_j$ and $b_j$ with
\begin{equation}
	\uuhh = \sum_{j\geqslant0} \frac{a_j}{r^{j+1}} - \ln r \sum_{j\geqslant4} \frac{b_j}{r^{j+1}} \, ,
\end{equation}
where $a_0$ is the Newtonian term, $a_1$ is the 1PN term and so on. Similarly, for use in applications involving $u^T$ we also introduce the coefficients $\alpha_j$ and $\beta_j$ in the expansion of the SF contribution
\begin{equation}
	u^T_\text{SF} = \sum_{j\geqslant0} \frac{\alpha_j}{r^{j+1}}  - \ln r \sum_{j\geqslant4} \frac{\beta_j}{r^{j+1}} \, .
\end{equation}
These series allow for the possibility of logarithmic terms, which are known not to start before the 4PN order. We also concluded in Paper II that $(\ln{r})^2$ terms cannot arise before the 5.5PN order. Since we are computing a conservative effect, possible time-odd logarithmic squared contributions at the 5.5PN or 6.5PN orders do not contribute. But there is still the possibility for a conservative 7PN $(\ln{r})^2$ effect, probably originating from a tail modification of the dissipative 5.5PN $(\ln{r})^2$ term. However, we shall not permit for such a small effect in our fits. As discussed below in Sec.~\ref{secVID}, we already have problems distinguishing the 7PN linear $\ln{r}$ term from the 7PN non-logarithmic contribution.

The analytically determined values of the coefficients $a_0$, $a_1$, $a_2$, $a_3$, $b_4$, $b_5$ and $\alpha_0$, $\alpha_1$, $\alpha_2$, $\alpha_3$, $\beta_4$, $\beta_5$ computed in Ref.~\cite{Det08} and Papers I-II are reported in Table \ref{known}.

\begin{table*}[h]
	\begin{center} 		\begin{tabular}{c | c || c | c} 			 \hline\hline
			coeff. & value & coeff. & value \\
			\hline
			$a_0$ & $-2$ & $\alpha_0$ & $-1$ \\
			$a_1$ & $-1$ & $\alpha_1$ & $-2$ \\
			$a_2$ & $-\frac{7}{4}$ & $\alpha_2$ & $-5$ \\
			$a_3$ & $-\frac{1387}{24} + \frac{41}{16}\pi^2$ & $\alpha_3$ & $-\frac{121}{3}+\frac{41}{32}\pi^2$ \\
			$b_4$ & $-\frac{128}{5}$ & $\beta_4$ & $-\frac{64}{5}$ \\
			$b_5$ & $+\frac{5944}{105}$ & $\beta_5$ & $+\frac{956}{105}$ \\
			\hline\hline 		\end{tabular} 		\caption{The analytically determined PN coefficients for $\uuhh$ (left) and $u^T_\text{SF}$ (right).} \label{known} 	\end{center}
\end{table*}

\subsection{Consistency between analytically and numerically determined PN coefficients}
\label{secVIC}

 In this Section we investigate the use of our data for $\uuhh$ and the fitting procedures we have described above (and expanded upon in the beginning of Sec.~\ref{secVID}).  We will begin by fitting for enough of the other PN coefficients to be able to verify numerically the various coefficients $a_3$, $b_4$ and $b_5$ now known from PN analysis.

As a first step in this Section, we will complete the task that was begun in \cite{Det08}, namely, the numerical determination of the coefficient $a_3$ (and $\alpha_3$), this time taking fully into account the known logarithmic terms at 4PN and 5PN order. For illustrative purposes only, these results are given in Table \ref{a3fit}. We were able to obtain a fit with six undetermined parameters, and could include data from $r=700$ down to $r=35$.  Note that, with the inclusion of the $b_4$ and $b_5$ coefficients, the precision of our tabulated value for $a_3$ has increased by more than four orders of magnitude from Paper I, although our accuracy is still no better than about $2\Sigma$. Such a discrepancy is not uncommon. The uncertainty, $\Sigma$, reflects only how well the data in the given, finite range can be represented by a combination of the basis functions. It is not a measure of the quality of a coefficient when considered as a PN expansion parameter, which necessarily involves an $r\rightarrow+\infty$ limiting process.

\begin{table*}[h]
\begin{center}
\begin{tabular}{c | l | l | l | l}
\hline\hline
3PN coeff. & \hspace{0.05cm} Ref. \cite{Det08} & \hspace{0.35cm} Paper I & \hspace{0.5cm} Paper II & \hspace{0.4cm} PN (exact) \\
\hline
$a_3$  & $-32.34(6)$ & $-32.479(10)$ & $-32.5008069(7)$ & $-32.50080538\cdots$ \\
$\alpha_3$  & $-27.61(3)$ & $-27.677(5)$ & $-27.6879035(4)$ & $-27.68790269\cdots$ \\
\hline\hline
\end{tabular}
\caption{ The results of a numerical fit for a set of six coefficients that includes $a_3$, which is now known analytically\cite{BDLW10a}. This fit uses the known results for $b_4$ and $b_5$\cite{BDLW10b}, but not the known value of $a_3$.  Thus, it is \textit{not} the best-fit of our data possible. The uncertainty in the last digit or two is in parentheses. The range runs from $r=35$ to $r=700$, with 266 data points and a respectable $\chsq$ of 264.} \label{a3fit}
\end{center}
\end{table*}

Our next step is to include the known value for $a_3$ and to use our numerical data to estimate values for the $b_4$ and $b_5$ coefficients. Our best quality numerical result was obtained with five fitted parameters, over a range from $r=700$ down to only $r=65$, and is given in the first row of Table \ref{1stfit}.  Notice that while our $b_4$ is determined relatively precisely, it has only about $6\Sigma$ accuracy.  The higher order coefficient $b_5$ is more difficult to obtain and, at this point, it is very poorly determined, but we can use the known value of $b_4$ in order to improve the accuracy for $b_5$. These results are presented in Table \ref{1stfit}, which again shows that we needed to fit for a total of six parameters to get a result of reasonable accuracy.  With this, we have reached a limit for treating our data in this way, since adding further parameters and inner points does not result in any higher quality fit.

\begin{table*}[h]
	\begin{center}
		\begin{tabular}{c | l | l | l | l | l | l | l | l}
			\hline\hline
			$r_\text{min}$& $\chi^2\!/$dof& \hspace{0.7cm} $a_4$ & \hspace{0.6cm} $b_4$ & \hspace{0.4cm} $a_5$ & \hspace{0.3cm} $b_5$ & \hspace{0.5cm} $a_6$ & \hspace{0.3cm} $b_6$ & \hspace{0.6cm} $a_7$ \\
			\hline
  			65 & \,\,0.961 & $-121.40(1)$ & $-25.612(2)$ & $-102(1)$ & $45.5(3)$ & $-2081(9)$ &\\ %% onerow5L7N3-a4.dat
			85 & \,\,0.976 & $-121.3180(7)$ & & $-91.5(7)$ & $48.5(2)$ & $-2170(8)$ &\\ %% onerow4L9N4-a4.dat
			65 & \,\,0.961 & $-121.313(1)$  & & $-79(2)$ & $50.6(4)$ & $-1868(44)$ & $131(21)$ \\ %% onerow5L9N4-a4.dat
			40 & \,\,0.969 & $-121.3052(6)$ & & $-47(1)$ & $55.7(2)$ & $-359(41)$ & $625(15)$ & $-7722(162)$ \\ %%onerow6L9N4-a4.dat
			\hline\hline
		\end{tabular}
		\caption{The numerically determined PN coefficients for $\uuhh$. Each row represents a different fit. The first two columns give the starting point $r_\text{min}$ at the inner boundary of the fitting range, and the value of $\chi^2$ statistic per degree of freedom (dof) for the chosen fit.  The degrees of freedom, $N-M$, for the fit, range between 212 and 255, depending on $r_\text{min}$.  If a value for a coefficient is not shown, then either that parameter was not included in that particular fit (far right) or its analytically known value was used (e.g., $b_4$). The formal uncertainty of a coefficient in the last digit or two is in parentheses. The outer boundary is at $700$ in each case.} \label{1stfit}
	\end{center}
\end{table*}

By now we have presented enough to show that we have data which allows high precision, with an accuracy that we now have some experience in relating to the computed error estimates. This experience will be valuable when we come to discuss further results in the next Section. For convenience, we summarize the relevant information further, in Table \ref{analytic}, referring just to our estimates of known PN parameters, and relating our error estimates to the observed accuracy.

\begin{table*}[h]
\begin{center}
\begin{tabular}{l | c | l | l | l }
\hline\hline
source & coeff. & \hspace{0.5cm} estimate  & accuracy & \hspace{0.3cm} exact result\\
\hline Paper I           & $\alpha_3$   & $-27.677(5)$     & $\rightarrow(11) $ &
$-27.6879\cdots$    \\
Table \ref{a3fit} & $a_3$        & $-32.5008069(7)$ & $\rightarrow(15) $ &
$-32.50080538\cdots$ \\
Table \ref{1stfit} & $b_4$        & $-25.612(2)$    & $\rightarrow(12)$ &
$-25.6$             \\
Table \ref{1stfit} & $b_5$        & $+55.7(2)$        & $\rightarrow(9)  $ &
$+56.6095\cdots$    \\
\hline\hline
\end{tabular}
\caption{Comparing the analytically known PN coefficients (column 5) with their numerically determined counterparts (column 3), and comparing the numerically determined error estimates (column 3) with the apparent accuracy (column 4). The source of the data is given in column 1.} \label{analytic}
\end{center}
\end{table*}

\subsection{Determining higher order PN terms numerically}\label{secVID}

In this Section we make maximum use of the coefficients which are already known. We find that in our \textit{best fit} analysis we can use a set of five basis functions corresponding to the unknown coefficients $a_4$, $a_5$, $a_6$, $b_6$ and $a_7$.

\begin{figure}
    \includegraphics[clip,angle=0,width=11.34cm,]{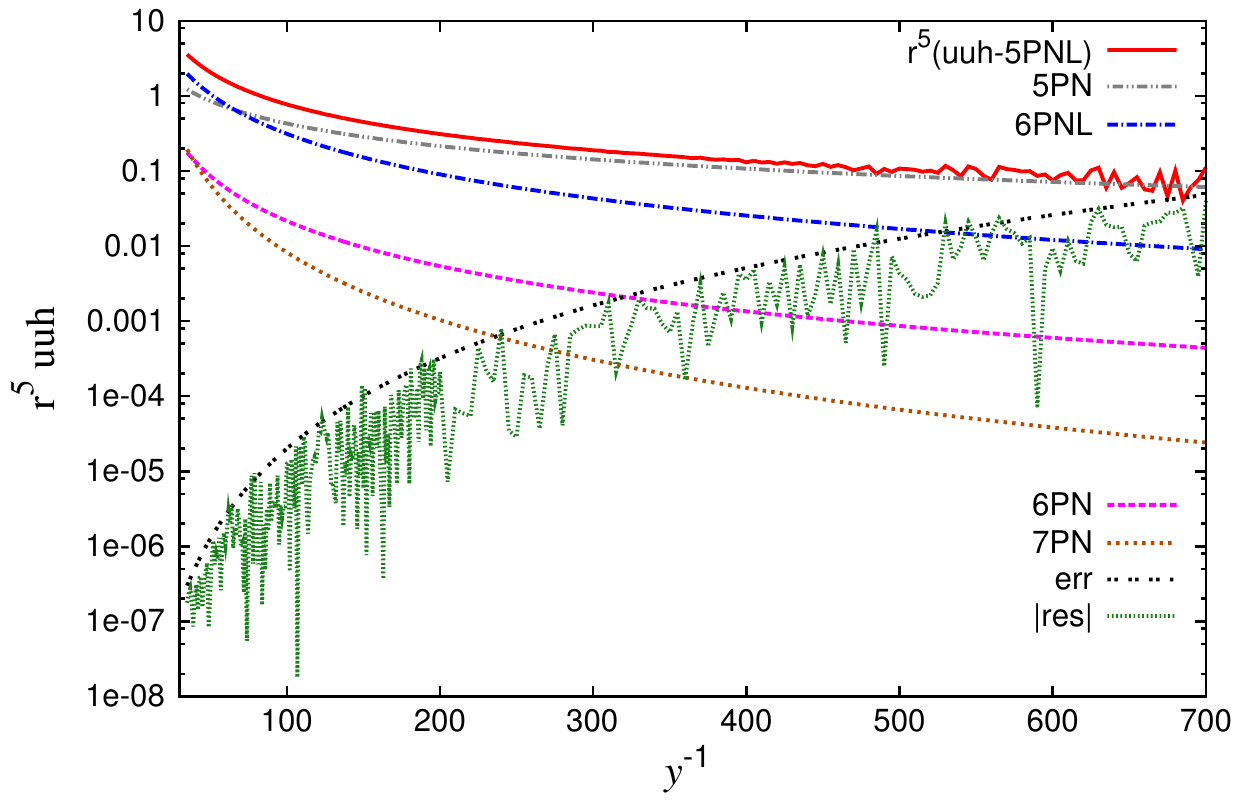}
\caption{The absolute value of the contributions of the numerically determined post-Newtonian terms to $r^5\uuhh$. Here PNL refers to just the logarithm term at the specified order. The contribution of $a_4$ is not shown but would be a horizontal line (since the 4PN terms behaves like $r^{-5}$) at approximately 121.3\,. The remainder after $a_4$ and all the known coefficients are removed from $r^5\uuhh$ is the top (red) continuous line. The lower (black) dotted line labelled ``err'' shows the uncertainty in $r^5\uuhh$, namely $2{\rm E} \, r^4 \times10^{-13}$. The jagged (green) line labelled ``$|$res$|$'' is the absolute remainder after all of the fitted terms have been removed. The figure reveals that, with regard to the uncertainty of the calculated $\uuhh$, the choice $E\simeq 1$ was slightly too large.}
\label{bestfig}
\end{figure}

\begin{table*}[h]
	\begin{center} 		\begin{tabular}{c | l || c | l} 			 \hline\hline
			coeff. & \hspace{0.5cm} value & coeff. & \hspace{0.5cm} value \\
			\hline
			$a_4$ & $-121.30310(10)$ & $\alpha_4$ & $-114.34747(5)$ \\
			$a_5$ & $-42.89(2)$      & $\alpha_5$ & $-245.53(1)$ \\
			$a_6$ & $-215(4)$        & $\alpha_6$ & $-695(2)$ \\
			$b_6$ & $+680(1)$        & $\beta_6$  & $+339.3(5)$ \\
			$a_7$ & $-8279(25)$      & $\alpha_7$ & $-5837(16)$ \\
			\hline\hline 		\end{tabular} 		\caption{The numerically determined values of higher-order PN coefficients for $\uuhh$ (left) and for $u_\text{SF}^T$ (right). The uncertainty in the last digit or two is in parentheses. The range runs from $r=40$ to $r=700$, with 261 data points being fit. The $\chsq$ statistic is 259. We believe that a contribution from a $b_7$ confounds the $a_7$ coefficient. Both terms fall off rapidly and have influence over the fit only at small $r$. And the radial dependence of these two terms only differ by a factor of $\ln r$ [or possibly $(\ln r)^2$] which changes slowly over their limited range of significance.} \label{bestfit} 	\end{center}
\end{table*}
In Table \ref{bestfit}, we describe the numerical fit of our data over a range in $r$ from $40$ to $700$. The $\chsq$ statistic is 259 and slightly larger than the degrees of freedom, 256, which denotes a good fit. Further, we expect that a good fit would be insensitive to changes in the boundaries of the range of data being fit, and we find, indeed, that if the outer boundary of the range decreases to $300$ then essentially none of the data in the Table changes, except for $\chsq$ and the degrees of freedom which decrease in a consistent fashion.  Figure \ref{bestfig} shows that in the outer part of the range $\uuhh$ is heavily dominated by only a few lower order terms in the PN expansion --- those above the lower black double-dashed line in the figure.

The inner edge of the range is more troublesome. The importance of a given higher order PN term decreases rapidly with increasing $r$. Moving the inner boundary of the range outward might move a currently well determined term into insignificance. This could actually lead to a smaller $\chsq$, but it would also lead to an increase in the $\Sigma_j$ of every coefficient. Moving the inner edge of the range inward might require that an additional higher order term be added to the fit. This extra term loses significance quickly with increasing $r$ so the new coefficient will be poorly determined and also result in an overall looser fit with an increase of $\Sigma_j$ for all of the coefficients. If the inner boundary and the set of basis functions are chosen properly, then a robust fit is revealed when the parameters being fit are insensitive to modest changes in the boundaries of the range. The fit described in Table \ref{bestfit} appears to be robust. The parameters in this Table are consistent with all fits with the inner boundary of the range varying from 35 to 45 and the outer boundary varying from 300 to 700.

If an additional term, with coefficient $b_7$, is added to the basis functions then, for identical ranges, each of the $\Sigma_j$ increases by a factor of about ten, and the changes in $a_4$ and $a_5$ are within this uncertainty. The coefficient $a_6$ changes sign and $b_6$ and $a_7$ change by an amount significantly larger than the corresponding $\Sigma_j$. And the new coefficient $b_7$ is quite large. In the context of fitting data to a set of basis functions these are recognized symptoms of over-fitting and imply that the extra coefficient degrades the fit.

\subsection{Summary}\label{secVIE}

\begin{figure}
    \includegraphics[width=7.88cm,angle=-90]{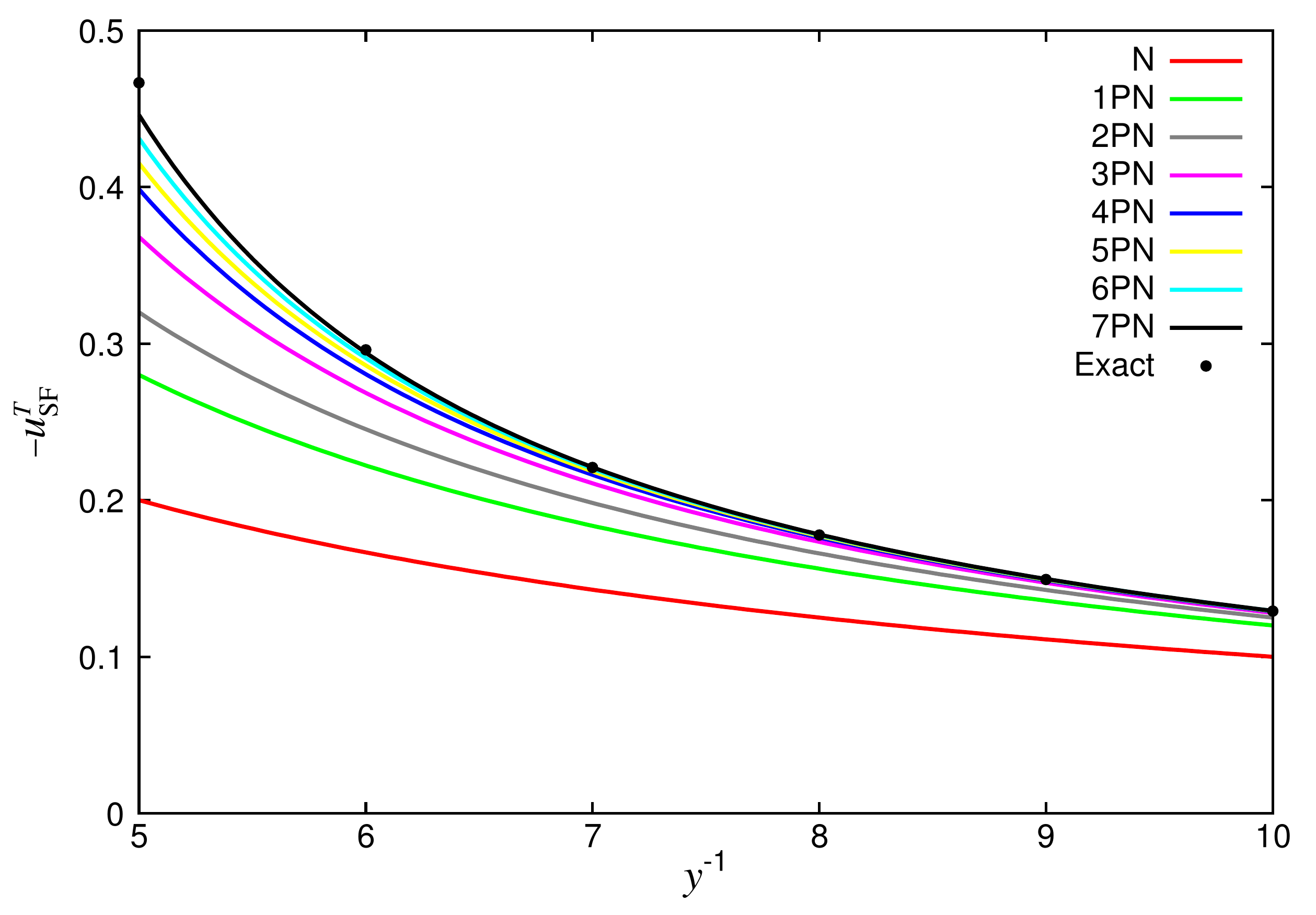}
    \caption{\footnotesize The self-force contribution $u^T_\mathrm{SF}$ to the redshift observable $u^T$, plotted as a function of the gauge invariant variable $y$. Note that $y^{-1}$ is an invariant measure of the orbital radius scaled by the black hole mass $m_2$ [see Eqs.~\eqref{ROmega} and \eqref{y}]. The ``exact'' numerical points are taken from Ref.~\cite{Det08}. Here, PN refers to all terms, including logarithms, up to the specified order (however recall that we did not include in our fit a log-term at 7PN order).}
    \label{uT_SF}
\end{figure}

Our best fit can be visualized in Fig.~\ref{uT_SF}, where we plot the self-force effect $u_\text{SF}^T$ on the redshift variable $u^T$ as a function of $r=y^{-1}$, as well as several truncated PN series up to 7PN order, based on the analytically determined coefficients summarized in Table \ref{known}, as well as our best fit of the higher-order PN coefficients reported in Table \ref{bestfit}. Observe in particular the smooth convergence of the successive PN approximations towards the exact SF results. Note, though, that there is still a small separation between the 7PN curve and the exact data in the very relativistic regime shown at the extreme left of Fig.~\ref{uT_SF}.

We have found that our data in the limited range of $35 \leqslant r \leqslant 700$ can be extremely well characterized by a fit with five appropriately chosen (basis) functions. That is, the coefficients in Table \ref{bestfit} are well determined, with small uncertainties, and small changes in the actual details of the fit result in coefficients lying within their error estimates. Fewer coefficients would result in a very poor characterization of the same data while more coefficients result in large uncertainties in the estimated coefficients, which themselves become overly sensitive to small changes in specific details (such as the actual choice of points to be fitted). In practice, over the data range we finally choose, and with the five coefficients we fit for, we end up with exceedingly good results for the estimated coefficients, and with residuals which sink to the level of our noise. We have a very high quality fit which is quite insensitive to minor details. Nevertheless, as Table \ref{analytic} hints, error estimates for these highest order coefficients should be regarded with an appropriate degree of caution.

\begin{acknowledgement} 
The authors acknowledge the 2008 Summer School on Mass and Motion, organized by A.~Spallicci and supported by the University of Orl\'eans and the CNRS, through which we experienced an extensive opportunity to understand each other's perspective and make rapid progress on this work.  SD and BFW acknowledge support through grants PHY-0555484 and PHY-0855503 from the National Science Foundation. LB and ALT acknowledge support from the Programme International de Coop\'eration Scientifique (CNRS--PICS). 
\end{acknowledgement}

\bibliography{}

\end{document}